\documentclass[aps,twocolumn,superscriptaddress,prl]{revtex4}
\usepackage{epsfig}
\usepackage{color}
\newcommand{\ben}{\begin{eqnarray}}
\newcommand{\een}{\end{eqnarray}}
\newcommand{\nnu}{\nonumber\\}
\newcommand{\bef}{\begin{figure}[htb]\centering}
\newcommand{\eef}{\end{figure}}
\newcommand{\sla}[1]{{#1}\!\!\!\slash}
\newcommand{\vacl}{\langle 0|}
\newcommand{\vacr}{|0\rangle}
\newcommand{\that}{\hat{T}}
\newcommand{\nbar}{\bar{n}}
\newcommand{\zhat}{\hat{z}}

\begin{document}
\title{Test of the Universality of Naive-time-reversal-odd
Fragmentation Functions}

\author{Dani\"el Boer}
\affiliation{Theory Group, KVI,
                University of Groningen,
                Zernikelaan 25, NL-9747 AA Groningen, The Netherlands}

\author{Zhong-Bo Kang}
\affiliation{RIKEN BNL Research Center,
             Brookhaven National Laboratory,
             Upton, NY 11973, USA}

\author{Werner Vogelsang}
\affiliation{Institute for Theoretical Physics,
                Universit\"{a}t T\"{u}bingen,
                Auf der Morgenstelle 14,
                D-72076 T\"{u}bingen, Germany}

\author{Feng Yuan}
\affiliation{Nuclear Science Division,
                Lawrence Berkeley National Laboratory,
                Berkeley, CA 94720, USA}
\affiliation{RIKEN BNL Research Center,
                Brookhaven National Laboratory,
                Upton, NY 11973, USA}

\begin{abstract}
  We investigate the ``spontaneous'' hyperon transverse polarization in
  $e^+e^-$ annihilation and semi-inclusive deep inelastic scattering
  processes as a test of the universality of the
  naive-time-reversal-odd transverse momentum dependent fragmentation
  functions. We find that 
  universality implies definite sign relations 
  among various observables. This provides a unique opportunity to study
  initial/final state interaction effects in the fragmentation process
  and test the associated factorization.
\end{abstract}

\date{\today}
\maketitle

The universality of parton distribution functions (PDFs) and fragmentation
functions (FFs) is a cornerstone of perturbative QCD applications
in hadronic physics. It relies on the factorization theorem
for the relevant high energy processes, which has been
rigorously proven and widely applied in phenomenological
studies. In particular, the PDFs depending only
on the longitudinal momentum fractions of hadrons carried by
the partons have been shown to be universal among different
processes~\cite{ColSopSte89}.
In recent years, however, intensive theoretical
investigations have shown that the so-called
naive-time-reversal-odd (T-odd) transverse momentum dependent (TMD)
PDFs are not universal due to differences
in initial/final state interactions.
For example, it was found that the difference
between the final state interactions in semi-inclusive
hadron production in deep inelastic scattering (SIDIS) and
initial state interactions in Drell-Yan lepton pair production (DY)
in hadronic collisions leads to an opposite sign in the
TMD Sivers function for quarks that enters in these two
processes~\cite{BroHwaSch02,Col02,BelJiYua02,BoeMulPij03,Siv90}.
Measurements of Sivers asymmetries in SIDIS have
been carried out experimentally, while those in the DY process are planned.
It is of great importance to test this modified universality
(sign change), in order to unravel fundamental
dynamics in strong interaction physics and the novel aspects of
nucleon structure involved in these processes.

On the other hand, their fragmentation counterparts, the T-odd TMD
FFs, have been shown to be
universal~\cite{Metz:2002iz,collins-s,Gamberg:2008yt,Meissner:2008yf,yuan-zhou}.
Here, the so-called Collins FFs~\cite{Col93} and polarizing
FFs \cite{Mulders:1995dh} have been the main focus.
The test of this universality (i.e., no sign change) is as important
as that for the modified
universality of the quark Sivers function discussed above.
The universality of these FFs involves the same QCD dynamics
and arguments for the TMD factorization of the relevant processes.

However, it is difficult to carry out a
test of the universality of the Collins FF because
of its chiral-odd nature. In order
to observe its effect, it has to be coupled with another
chiral-odd object. For example, in the SIDIS process,
the Collins function couples to the quark transversity
distribution to generate a novel single transverse
spin asymmetry; or in the $e^+e^-$ annihilation process,
two Collins functions couple together and lead to an
asymmetric azimuthal angular correlation of di-hadron
production. The issue is that the sign of these chiral-odd
objects is not easy to determine. For example, one can
change the sign of all functions involved without altering
the actual physical observables. Thus, in order to
test universality, one needs to be able to determine the
relative signs, which will necessarily involve additional
chiral-odd functions and therefore further observables.

In this paper, we investigate a test of the universality
of the T-odd TMD FFs by studying hyperon polarization
in the SIDIS and $e^+e^-$ annihilation
processes. The relevant T-odd FFs
describe the transverse polarization of the hyperon, typically a
$\Lambda$, correlated
with its transverse momentum relative to the fragmenting
quark jet~\cite{Mulders:1995dh,Boer:1997mf,Anselmino:2000vs,Anselmino:2001js}.
The universality of this fragmentation function
was first demonstrated in a model calculation in
Ref.~\cite{Metz:2002iz}, and subsequently using
model-independent arguments~\cite{Meissner:2008yf}.
It is a chiral-even function, and
will couple to chiral-even functions in hadronic processes,
such as the spin-averaged quark PDFs in SIDIS and the
spin-averaged FFs in $e^+e^-$ processes.
Because these functions are well determined
and most importantly all positive,
it will be possible to unambiguously
measure the sign of the T-odd FFs and test their universality.

The TMD quark FFs
can be defined through the following correlator:
\ben
\Delta(z_h, p_\perp)&=&\frac{1}{z_h}\int\frac{dy^-d^2y_\perp}{(2\pi)^3}e^{ik\cdot y}
\vacl {\cal L}_y\psi(y)|P_h S_\perp X\rangle
\nnu
&&\,
\times\langle P_h S_\perp X| \bar{\psi}(0){\cal L}_0^\dagger\vacr|_{y^+=0},
\label{ff}
\een
where $P_h$ is the momentum of the final state hadron with spin $S_\perp$,
which has a
transverse component $p_\perp$ relative to the momentum $k$ of the
fragmenting quark.
We choose the hadron to move along the $+z$ direction, and
define the light-cone components $p^\pm=(p^0\pm p^z)/\sqrt{2}$.
For convenience, we define two light-like vectors:
$\nbar^\mu=\delta^{\mu+}$ and
$n^\mu=\delta^{\mu-}$.
The momentum fraction $z_h=P_h^+/k^+$, and
$\vec{k}_\perp=-\vec{p}_\perp/z_h$.
The correlator can then be expanded as
\ben
\Delta(z_h, p_\perp)&=&\frac{1}{2}\left[D(z_h, p_\perp^2)
\sla{\nbar}
+\frac{1}{M_h}D_{1T}^{\perp}(z_h, p_\perp^2)
\right.
\nnu
&&\,\left.
\times
\epsilon^{\mu\nu\rho\sigma}\gamma_\mu\nbar_\nu p_{\perp\rho}S_{\perp\sigma}+\cdots\right],
\een
where we have only kept the FFs of relevance here: the spin-averaged
$D(z_h,p_\perp^2)$, and the spin-dependent T-odd one
$D_{1T}^\perp(z_h,p_\perp^2)$, usually called ``polarizing FF''. The latter
leads to single-transverse-spin
asymmetries in hyperon production in various
processes~\cite{Boer:1997mf,Anselmino:2000vs,Anselmino:2001js}.
It offers an explanation for the ``spontaneous'' hyperon polarization
observed in hadron-hadron collisions many years
ago~\cite{hyper}. 

In Eq.~(\ref{ff}), the gauge link is defined as
${\cal L}_y={\cal P}\exp\left(ig\int_0^\infty d\lambda v\cdot A(y+\lambda v)\right)$
with $v$ satisfying $v^-\gg v^+$ and $v^2\neq 0$
to regulate the light-cone
singularity~\cite{BelJiYua02}.
The gauge link results from the initial/final state interactions.

The polarizing FF cannot be calculated from first principles, but it
can be computed in perturbation theory for large
$p_\perp$, yielding its so-called perturbative tail.
Since this has not been given elsewhere we present it here:
\ben
D_{1T}^\perp(z_h,p_\perp^2) & = & \frac{\alpha_s}{2\pi^2}
\frac{M_h}{(p_\perp^2)^2} \int \frac{dz}{z}
\bigg[ A(z)\nnu
& & \mbox{}\hspace{-1 mm} +\delta(\hat{z}-1) \hat{T}(z) C_F \left(\ln
    \frac{\hat{\zeta^2}}{p_\perp^2}-1\right) \bigg],
\label{DT}
\een
where $\hat{z}=z_h/z, \hat\zeta^2=(v\cdot P_h)^2/v^2$, and $A$
is given by:
\ben
A(z)&=&C_F\left[-(1+\zhat^2)z\frac{\partial \that(z)}{\partial z}-
\that(z)\frac{2\zhat^3-3\zhat^2-1}{(1-\zhat)_+}\right]
\nnu
&&
+\int\frac{dz_1}{z_1^2}{\rm
  PV}\left(\frac{1}{\frac{1}{z}-\frac{1}{z_1}}\right)\that_F(z,
z_1)\nnu
&& \times
\left[C_F\left(\frac{z}{z_1}-\frac{z_h}{z_1}+\frac{z_h}{z}-\frac{z_h^2}{zz_1}-2\right)
\right.
\nnu
&&
\left.
+\frac{C_A}{2}\frac{(zz_h+z_1z_h-2zz_1)(zz_1+z_h^2)}{(z-z_1)(z_1-z_h)z^2}\right] \ .
\label{AA}
\een
We note that the first two terms in $A$ have the same structure as
the soft-pole contributions to the
perturbative tail of the Sivers PDF~\cite{Ji:2006ub}.
In Eqs.~(\ref{DT}),(\ref{AA}) we have
\ben
\that(z)&=&z^2\int \frac{dy^-}{2\pi} e^{ik^+ y^-} \frac{1}{2}
\left\{{\rm Tr} \,\sla{n}\vacl \epsilon^{n\nbar S_\perp\alpha}
\left[iD_{\perp\alpha}
\right.\right.
\nnu
&&\,\left.\left.
+\int_{y^-}^{+\infty} d\xi^- gF_{\alpha}^{~+}(\xi^-)\right]
\psi(y^-)|P_hS_\perp X\rangle\right. \nonumber\\
&&\left.\times \langle P_hS_\perp X| \bar{\psi}(0)\vacr+{\rm h.c.}\right\} \ ,
\een
with $D_\perp^\alpha=\partial_\perp^\alpha-igA_\perp^\alpha$
the covariant derivative, $F^{\alpha\beta}$ the field strength tensor,
and
\ben
\that_F(z_1, z_2)&=&z_1z_2\int \frac{dy^-}{2\pi} \frac{d\xi^-}{2\pi}
e^{ik_2^+ y^-}e^{ik_g^+\xi^-}
\nnu
&&\,\hspace*{-8mm}\times\frac{1}{2}
 \left\{{\rm Tr} \,\sla{n}
\vacl \epsilon^{n\nbar S_\perp\alpha} gF^{+}_{\;\;\;\alpha}(\xi^-)
\psi(y^-)|P_hS_\perp X\rangle\right. \nonumber\\
&&\,\hspace*{-8mm}
\left.\times \langle P_hS_\perp X| \bar{\psi}(0)\vacr+{\rm h.c.}\right\} \ ,
\een
where $k_g^+=k_1^+-k_2^+$ with $k_1^+=P_h^+/z_1$ and
$k_2^+=P_h^+/z_2$. The twist-three correlator $\that$ is related to
$D_{1T}^\perp$ as follows:
\ben
\that(z)& = & \int d^2 p_\perp
\frac{|\vec{p}_\perp|^2}{M_h}D_{1T}^{\perp}(z, p_\perp^2).
\een
In the calculation of the perturbative tail we have verified
explicitly the gauge-link independence of the result, which is
equivalent to establishing the
universality of $D_{1T}^\perp$ in the perturbative region.
Arguments for its general universality have been given
based on the vanishing of
the so-called gluonic pole matrix elements in the
fragmentation function~\cite{Gamberg:2008yt,Meissner:2008yf}.
The universality is based on the property~\cite{Meissner:2008yf}
$\that_F(z_1,z_2)|_{z_1\ge z_2}=0$, or equivalently,
$\that_F(z_1,z_2)|_{k_g^+\le 0}=0$. Intuitively, this means
that the parton momentum entering hadronic matrix elements
in the fragmentation function has to be positive.

This property also turns out to be essential for the universality
of the perturbative tail, and our calculation hence provides a consistency
check for the arguments in Refs.~\cite{Gamberg:2008yt,Meissner:2008yf}.
\bef
\psfig{file=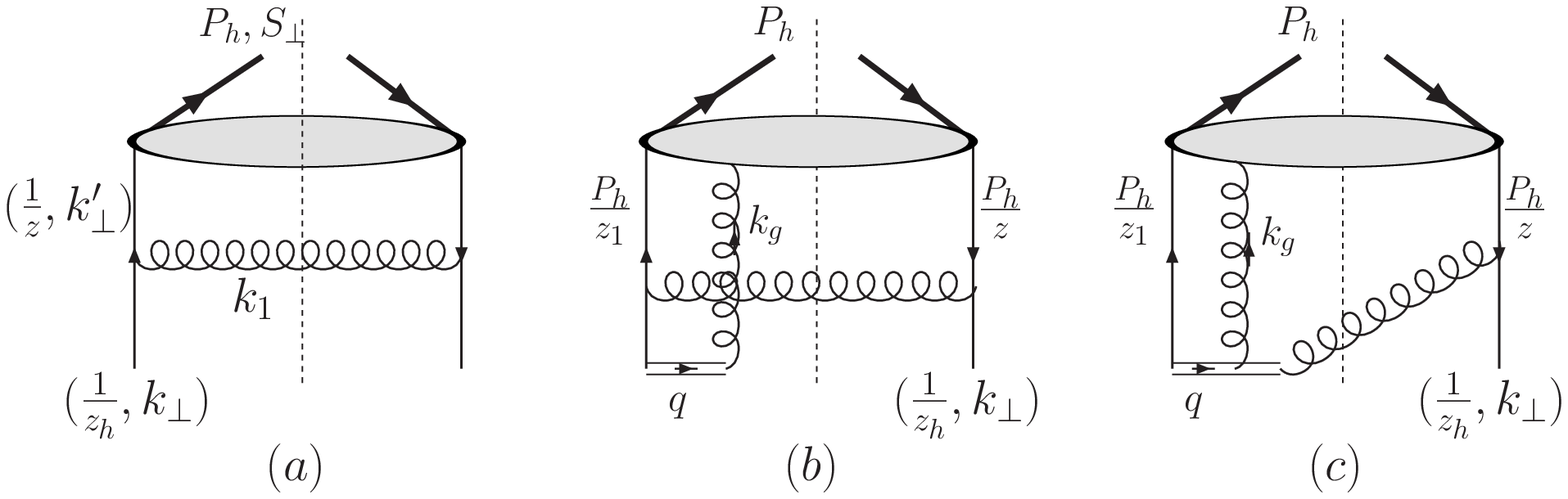, width=0.45\textwidth}
\caption{\it Sample Feynman diagrams contributing
to the naive T-odd spin dependent FF
at large transverse momentum. The double lines in (b,c) represent
the eikonal propagators from the gauge link expansion of
the fragmentation function definition. The pole contributions
from the eikonal propagators vanish.}
\label{f1}
\eef
At large transverse momentum,
a hard gluon has to be radiated, as shown in
Fig.~\ref{f1}: (a) is a typical quark splitting contribution to the
fragmentation function;
(b) and (c) are sample diagrams with gauge link contributions.
The gauge link represents the initial/final state interactions
in the associated processes. In the calculation of the gauge link,
eikonal propagators of the form $1/(q^+\pm i\epsilon)$ arise,
where the $i\epsilon$ prescription depends on the gauge link
direction (initial or final state interaction). As is
well-known,
this is responsible for the nontrivial universality properties of the
T-odd Sivers functions.
For the FFs, however, the pole contributions
from the gauge link vanish, such that the T-odd FFs
are universal among different
processes~\cite{Metz:2002iz,collins-s,Meissner:2008yf}.
Specifically, in the diagram in Fig.~\ref{f1}(b)
the pole contribution (soft pole) from the eikonal propagator
leads to a delta-function
$\delta(q^+)=\delta(k_g^+)$, where the associated
twist-three correlation function $\that_F(z,z_1)|_{k_g^+=0}$
vanishes~\cite{Meissner:2008yf}.
Similarly, in Fig.~\ref{f1}(c), the pole contribution (hard pole) leads
to $\delta(q^+)=\delta(k_g^++k_1^+)$ where $k_1$ is the momentum of the
radiated gluon. Since $k_1^+>0$ by kinematics which results
in $k_g^+<0$, the associated twist-three function vanishes again,
$\that_F(z,z_1)|_{k_g^+<0}=0$. All other diagrams
can be analyzed accordingly. Therefore, there is no contribution
from either the soft or hard poles associated with the eikonal
propagator in the gauge link expansion. The conclusion is that
the T-odd fragmentation function calculated from these
perturbative diagrams does not depend on the gauge link direction,
and hence is universal.
We emphasize, however, that the gauge link contributions are
important for guaranteeing the gauge invariance of the
fragmentation function.

We now discuss the possibility to test the 
universality of $D_{1T}^\perp$ in experiment.
Here we consider the $\Lambda$ hyperon as an example. 
In our calculations, we focus on the polarization of $\Lambda$ only, 
but extension to $\bar{\Lambda}$ and $\Lambda+\bar{\Lambda}$ 
can follow similarly.
As it turns out, the comparison of the
asymmetries induced by $D_{1T}^\perp$ in SIDIS ($\ell p \to \ell'
\Lambda^\uparrow X$) \cite{Mulders:1995dh,Anselmino:2001js}
and $e^+ e^-$ annihilation ($e^+ e^- \to
\Lambda^\uparrow {\rm jet} X$) \cite{Boer:1997mf} is not straightforward. The
problem is the possibility of cancellation among different flavors,
which depends on the unknown magnitude of $SU(3)$ breaking and which can lead
to sign reversal of the asymmetries. This also applies to another test
of universality of $D_{1T}^\perp$ proposed in Ref.\ \cite{Boer:2007nh}.
Here we will show that detection of the $\Lambda$ 
in coincidence with a
light hadron $h$ in $e^+ e^-$ annihilation alleviates
the situation significantly and offers a robust test of the universality,
to be performed for instance with SIDIS at COMPASS, HERMES, 
Jefferson Lab, or a future
electron-ion collider (EIC) and with $e^+ e^-$ at BELLE.

The hyperon (transverse) polarization is defined as
$P_T^\Lambda=d\Delta\sigma(S_\perp)/d\sigma$
with $d\Delta\sigma(S_\perp)=\frac{1}{2}
\left[d\sigma(S_\perp)-d\sigma(-S_\perp)\right]$
and $d\sigma=\frac{1}{2}
\left[d\sigma(S_\perp)+d\sigma(-S_\perp)\right]$.
The differential cross section for a correlation of a
$\Lambda$ and a light hadron in $e^+e^-$,
$e^-(\ell)+e^+(\ell')\to \gamma^*(q)\to\Lambda^\uparrow(P_\Lambda, S_\perp)
+ h(P_h) +X$
can be written as
\ben
\frac{d\Delta\sigma(S_\perp)}{dz_\Lambda dz_h dy d^2 q_\perp}&=&
\sigma_0\epsilon^{\alpha\beta} S_{\perp\alpha}q_{\perp\beta}\frac{1}{q_{\perp}^2}
z_\Lambda^2 z_h^2 \int_\perp
\frac{-z_\Lambda\vec{k}_{\Lambda\perp} \cdot \vec{q}_{\perp}}{M_\Lambda}
\nnu
&&\times
D_{1T}^{\perp}(z_\Lambda, z_\Lambda^2k_{\Lambda\perp}^2)
D(z_h, z_hk_{h\perp}^2)
\nnu
&&\times
(S(\lambda_\perp))^{-1}H(Q^2) \ ,
\label{sidistmd}
\een
where $Q^2=q^2=(\ell+\ell')^2$, $z_i=2P_i\cdot q/Q^2$ with $i=\Lambda,h$,
$y=P_h\cdot \ell /P_h\cdot q$, $\sigma_0=N_c 4\pi\alpha_{em}^2(1/2-y+y^2)/Q^2$
with $\alpha_{em}$ the electromagnetic coupling, and
the integral symbol $\int_\perp \equiv
\int d^2k_{\Lambda\perp} d^2k_{h\perp} d^2\lambda_\perp\delta^2(\vec{k}_{\Lambda\perp}+\vec{k}_{h\perp}+\vec{\lambda}_\perp-\vec{q}_{\perp})$. $q_\perp$ is related to the ``transverse'' component of the virtual photon momentum
defined as
$q_t^\mu=q^\mu-\frac{P_h\cdot q}{P_h\cdot P_\Lambda}P_\Lambda^\mu
-\frac{P_\Lambda\cdot q}{P_\Lambda\cdot P_h}P_h^\mu$,
with $q_\perp^2=-q_t^\mu q_{t\mu}$.

Similarly, the differential cross section for $\Lambda$ production in SIDIS,
$e(\ell)+p(P)\to e (\ell')+\Lambda^\uparrow(P_\Lambda, S_\perp)+X$, reads
\ben
\frac{d\Delta\sigma(S_\perp)}{dx_B dy dz_\Lambda d^2 P_{\Lambda\perp}}&=&
\sigma_0^{DIS}\epsilon^{\alpha\beta} S_{\perp\alpha}P_{\Lambda\perp\beta}\frac{1}{P_{\Lambda\perp}^2}
\int_\perp
\frac{\vec{p}_\perp \cdot \vec{P}_{\Lambda\perp}}{M_\Lambda}
\nnu
&&\times
q(x_B, k_\perp)D_{1T}^{\perp}(z_\Lambda, p_\perp)
\nnu
&&\times
(S(\lambda_\perp))^{-1}H(Q^2)\ ,
\label{eetmd}
\een
where $Q^2=-q^2=-(\ell'-\ell)^2$,
$x_B=Q^2/2P\cdot q$, $z_\Lambda=P\cdot P_\Lambda/ P\cdot q$,
$y=P\cdot q/P\cdot \ell$, $\sigma_0^{DIS}=4\pi\alpha_{\rm em}^2(1/y-1+y/2)/Q^2$,
$q(x_B,k_\perp)$ is the spin-averaged quark distribution,
and $\int_\perp\equiv\int d^2k_\perp d^2p_{\perp} d^2\lambda_\perp\delta^2(z_\Lambda\vec{k}_\perp+\vec{p}_\perp+\vec{\lambda}_\perp-\vec{P}_{\Lambda\perp})$.
In both Eqs.~(\ref{sidistmd}) and (\ref{eetmd}), $S(\lambda_\perp)$
and $H(Q^2)$ denote soft and hard factors, respectively.
We have ignored contributions from chiral-odd
functions~\cite{Boer:1997nt,Kanazawa:2000mt,Zhou:2008fb} that lead to
different azimuthal dependences and therefore can be either distinguished
or averaged out.

We will now estimate the $\Lambda$ polarization $P_T^\Lambda$
in both processes.
We will assume the lowest order results for the soft and
hard factors in both Eqs.~(\ref{sidistmd}) and (\ref{eetmd}):
$(S(\lambda_\perp))^{-1}=\delta^2(\lambda_\perp)$ and
$H(Q^2)=1$. 
As in \cite{Anselmino:2000vs,Anselmino:2001js},
we keep the transverse momentum dependence
only in $D_{1T}^{\perp}$ and drop it in the spin-averaged PDFs
and FFs. We choose the initial hadron in SIDIS
or the final hadron in $e^+e^-$ along the $-z$ direction, while
$x$ is along the transverse momentum of the $\Lambda$.
Thus the transverse polarization is measured along $+y$.
The azimuthal angle dependence will be
$d\sigma(S_\perp)\propto |S_\perp| P_{\Lambda\perp} \sin\phi$,
where $\phi$ is the angle between $S_\perp$ and $P_{\Lambda\perp}$,
and $\phi=\pi/2$ in our frame.

We adopt the parametrizations of $D_{1T}^\perp$
obtained in \cite{Anselmino:2001js},
\ben
D_{1T}^{\perp q}(z_h, k_\perp)=f^\Delta(z_h, k_\perp)D_{\Lambda/q}(z_s, k_\perp),
\een
where
$D_{1T}^{\perp q}$ is proportional to the spin-averaged
$\Lambda$ FFs $D_{\Lambda/q}(z_h, k_\perp)$ with the proportionality
coefficient
$f^\Delta(z_h, k_\perp)$ a function fitted to the available
experimental data.
The $k_\perp$-dependence has been assumed to be Gaussian in both functions.
Several spin-averaged FFs have been considered, such as the $SU(3)$ flavor
symmetric \cite{deFlorian:1997zj} or broken \cite{Indumathi:1998am} ones.
Since both types of parametrizations fit the available data, a strong
dependence on $SU(3)$ breaking would jeopardize the test of
universality in the case of small asymmetries.
In Fig.~\ref{f2}, we show the $\Lambda$ polarization as a function
of $z_\Lambda$ in both semi-inclusive DIS $ep\to e'+\Lambda^\uparrow+X$
(left) and $e^+e^-$ annihilation
$e^+e^-\to \pi^\pm+\Lambda^\uparrow+X$ (right). For the latter, we have
used the pion fragmentation functions of Ref.~\cite{deFlorian:2007aj}.
\bef
\psfig{file=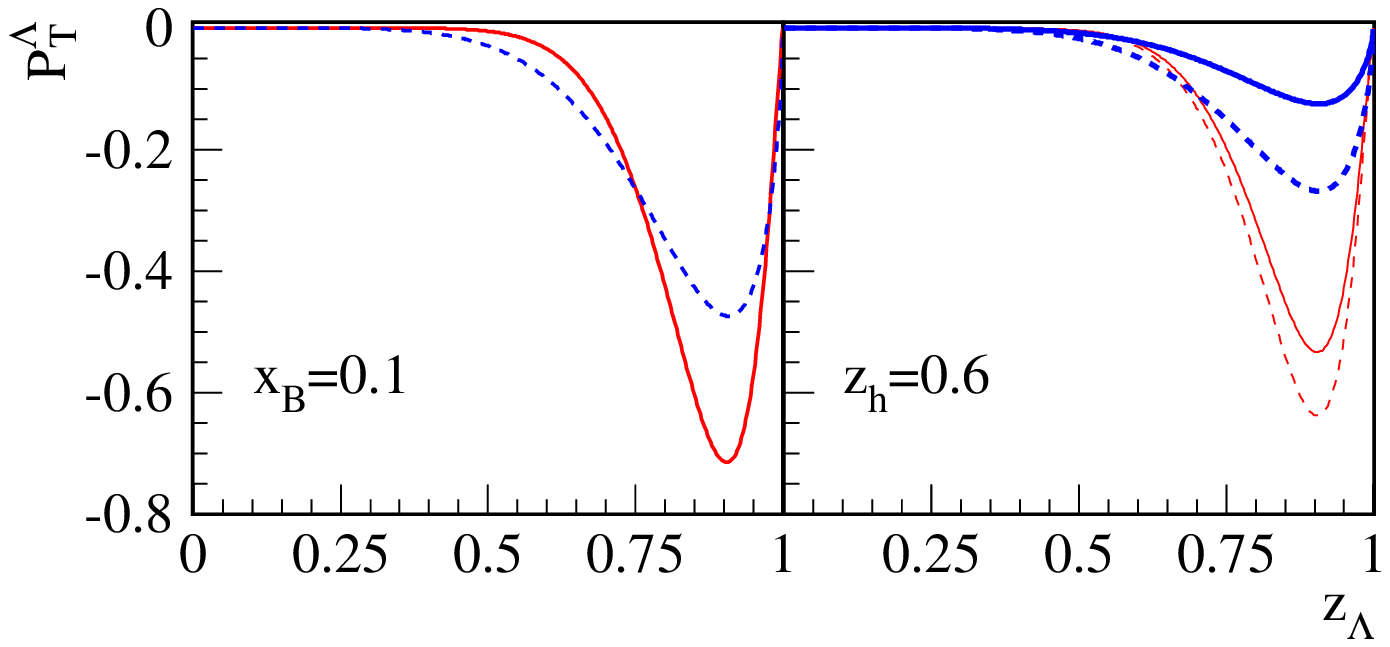, width=0.45\textwidth}
\caption{\it Universality predictions
of transverse $\Lambda$ polarization $P_T^\Lambda$
as function of $z_\Lambda$.
Left: SIDIS $ep\to e'+\Lambda^\uparrow+X$ at
$\langle Q^2\rangle \sim 10$ GeV$^2$ and $x_B\sim 0.1$. We have
integrated over $P_{\Lambda\perp}\leq 3$~GeV.
The solid (dashed) line is for the $SU(3)$-symmetric (broken)
spin-averaged $\Lambda$ FFs.
Right: $e^+e^-\to \pi^\pm+\Lambda^\uparrow+X$
at $\langle Q^2\rangle \sim 100$ GeV$^2$ and $z_h\sim 0.6$.
Thin (thick) lines are for the $SU(3)$-symmetric (broken)
spin-averaged $\Lambda$ FFs. The solid (dashed) lines
are for $\pi^+$ ($\pi^-$).}
\label{f2}
\eef
Our numerical results show that the polarizations are sizable,
and have the same sign in both processes, irrespective of the
size of the $SU(3)$ breaking. In both SIDIS and $e^+e^-$ annihilation,
up quarks dominate the scattering.
Since $D_{1T}^{\perp u}$ is negative in the models we are using,
the negative sign of $P_T^\Lambda$ in SIDIS follows immediately.
For $e^+e^-\to \pi^-+\Lambda^\uparrow+X$, the dominant mechanism
is up-quark fragmentation to the $\Lambda$ and $\bar{u}$-fragmentation
to the $\pi^-$, again resulting in a negative asymmetry. For
positive pions, there is some competition among the various flavor
combinations, and there is a larger contribution by $D_{1T}^{\perp d}$.
Since the latter is negative as well, the asymmetry remains negative.
Therefore universality implies equality of the signs of the asymmetries in
the two processes. 
Fig.~\ref{f3} shows that the process
$e^+e^-\to {\rm jet}+\Lambda^\uparrow+X$, on the other hand,
is much more sensitive to
flavor cancellation effects, and its sign relation to the SIDIS
polarization would not yield a robust test.
Also shown in Fig.~\ref{f3} is $\Lambda$+kaon production in
$e^+e^-$ annihilation. We observe
that $K^-$ production gives results very similar to those for
$\pi^-$, so that $\Lambda$ plus negatively charged hadrons would
likely do equally well. For positive kaons, we find that the $\Lambda$
polarization flips sign, as a result of the relatively bigger 
contribution from the $s\to\Lambda$ fragmentation function, 
which is expected positive. 
It will thus be possible to unambiguously measure the signs of the T-odd  
FFs. With precise experimental data one could further perform 
a global analysis to see if a universal set of polarizing  
FFs could be obtained. 
We note that the opposite signs of the $u/d$ and $s$ quark polarizing
FFs used here arise naturally 
to ensure that $\bar{\Lambda}$ polarization in $p p$ collisions
is consistent with zero \cite{Anselmino:2000vs}. 
Although the uncertainties in the polarizing FFs are 
rather large in general, as also witnessed by the fact that the
calculated polarization may exceed unity, 
the principle of fixing the sign 
by selecting $u/d$ or $s$ quark dominanted processes 
is expected to be robust. 
\bef
\psfig{file=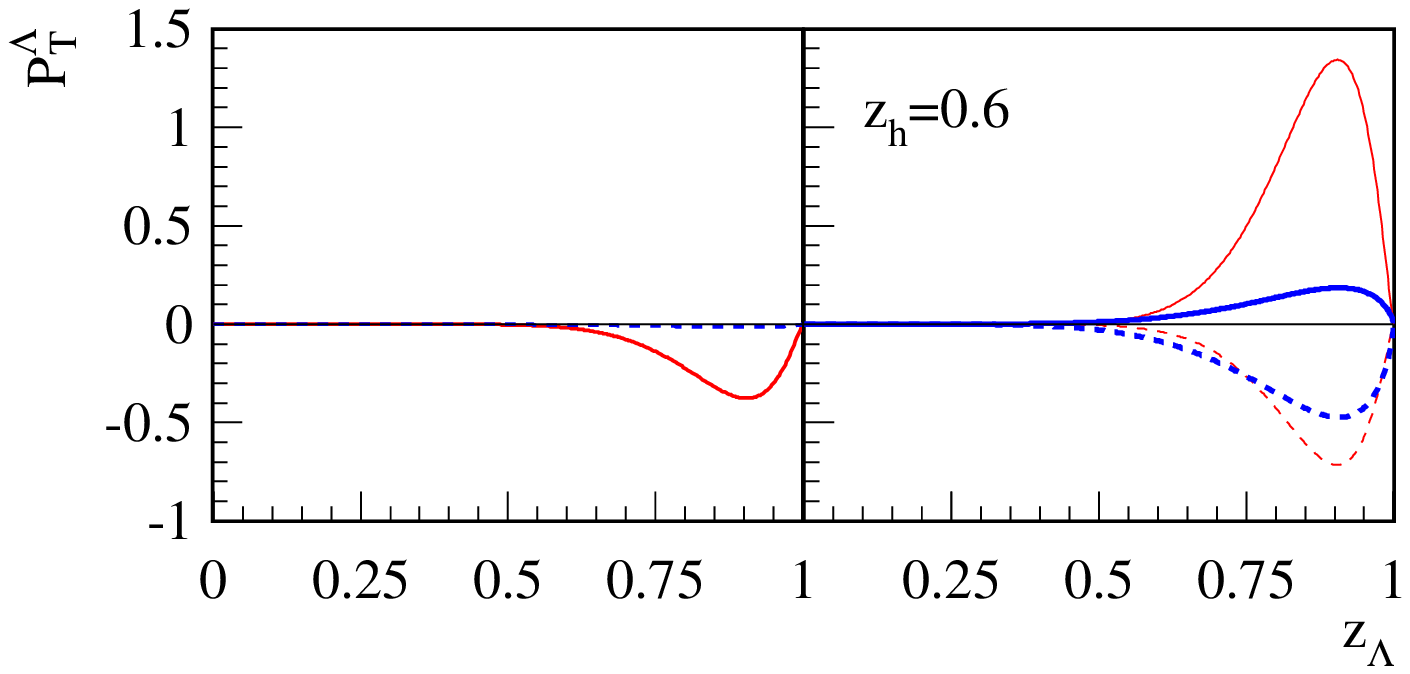, width=0.45\textwidth}
\caption{\it $\Lambda$ hyperon (transverse) polarization $P_T^\Lambda$
as function of $z_\Lambda$ at $\langle Q^2\rangle \sim 100$ GeV$^2$.
Left: $e^+e^-\to {\rm jet}+\Lambda^\uparrow+X$.
The solid (dashed) line is for the $SU(3)$-symmetric (broken)
spin-averaged $\Lambda$ FFs.
Right: $e^+e^-\to K^\pm+\Lambda^\uparrow+X$
at $z_h\sim 0.6$.
Thin (thick) lines are for the $SU(3)$-symmetric (broken)
spin-averaged FFs. The solid (dashed) lines
are for the $K^+$ ($K^-$).}
\label{f3}
\eef

In summary, we have investigated the universality of the
T-odd spin-dependent FFs using hyperon (transverse) polarization
in $e^+e^-$ annihilation and semi-inclusive DIS
processes. Definite signs for the polarization are predicted
based on the current knowledge of the polarizing fragmentation
functions. Despite the large uncertainties in these functions, 
the obtained sign relations among the measured polarizations
constitute a robust test of universality, 
in the sense that they are necessary conditions for universality 
to hold. We hope that this test can be carried out in the 
near future. 
It would provide an important confirmation of our current
understanding of novel single spin asymmetries in high energy
hadronic reactions and the associated QCD dynamics.

This work was supported in part by the
U.S. Department of Energy
under grant number DE-AC02-05CH11231 (FY) and
contract number DE-AC02-98CH10886 (ZK, FY and WV).


\end{document}